\documentclass[aps,prc,groupedaddress,showkeys,showpacs]{revtex4}
\usepackage{amssymb}

\usepackage{graphicx}

\begin{document}

\title{The influence of reconstruction criteria on the sensitive probes of the symmetry potential}

\author {Qingfeng Li$\, ^{1,2}$\footnote{e-mail address: liqf@fias.uni-frankfurt.de}
\email[]{Qi.Li@fias.uni-frankfurt.de}}
\address{
1) Frankfurt Institute for Advanced Studies (FIAS), Johann Wolfgang
Goethe-Universit\"{a}t, Max-von-Laue-Str.\ 1, D-60438 Frankfurt am
Main, Germany \\
2) School of Science, Huzhou Teachers College, Huzhou 313000, China
 }


\begin{abstract}
Different criteria of constructing clusters and tracing back $\Delta$
resonances from the intermediate-energy neutron-rich HICs are
discussed by employing the updated UrQMD transport model. It is
found that both the phase-space and the coordinate-density criteria affect
the single and the double neutron/proton ratios of free nucleons at small
transverse momenta, but the influence becomes invisible at large
transverse momenta. The effect of different methods of
reconstructing freeze-out $\Delta$s on the $\Delta^0/\Delta^{++}$
ratio is strong in a large kinetic energy region.
\end{abstract}

\keywords{Transport model; symmetry potential; sensitive probe; reconstruction.}

\pacs{24.10.Lx, 25.75.Dw, 25.75.-q}

 \maketitle

\section{Introduction}

The equation of state (EoS) of nuclear matter is one of the most
important topics in the field of low and intermediate energy heavy
ion collisions (HICs). In recent years, more and more work have been
focusing on the isospin asymmetry of the HICs, which
introduces more uncertainties into the EoS. Without considering the
momentum dependence of the EoS, the isospin dependent EoS for
asymmetric nuclear matter can be simply expressed as
$e(u,\delta)=e_0(u)+e_{sym}\delta^2$ where $u=\rho/\rho_0$ is the
reduced nuclear density and $\delta=(\rho_n-\rho_p)/\rho$ is the
isospin asymmetry in terms of neutron and proton densities. $e_0$
and $e_{sym}$ are the isospin independent term (which includes the
Skyrme and the Yukawa potentials) and the symmetry energy (which can
be expressed as $e_{sym}=S_0F(u)$ where $S_0$ is the symmetry energy
coefficient and F(u) the density dependence), respectively (see, e.g., Refs.\ \cite{BaoAnBook01,baranRP}).
Since the symmetry energy term plays also an important role in
nuclear structure and astrophysics, it indeed deserves more
attention. With the many efforts from both theoretical and
experimental sides, the uncertainties in EoS have been largely
constrained although arguments remain in this field. For example,
(1) the isospin-independent EoS has been constrained into a soft
compress modulus ($K\simeq 190-270$ MeV, see, for example,
\cite{Vretenar:2003qm,Fuchs:2007vt,Youngblood:1999aa,Danielewicz:2002pu}).
(2) The symmetry energy coefficient has been constrained into the
region $S_0 \simeq 30-36$ MeV
\cite{Vretenar:2003qm,Pomorski03,Klimkiewicz:2007zz}. And, (3) the strength
factor ($\gamma$) of the density dependence of the symmetry
potential is also shown not to be stiff at subnormal densities
\cite{Chen:2004si,Li:2005jy}, i.e., $\gamma =0.69 $ or $1.05$ when the form $F(u)=u^\gamma$ is employed,
which depends on the treatments of nucleon-nucleon collisions in the
BUU transport model.

While the density dependence of the symmetry potential at supranormal densities
is still quite uncertain. So far several
(probably) sensitive probes especially on this issue have been
brought out by theoretical groups, which are hoped to be observed
from experiments such as the planned FRIB(USA), the upcoming GSI new facility FAIR
(Germany), and the cooling storage ring
CSR being tested at Lanzhou (China), etc. For example, the $\pi^-$
to $\pi^+$ multiplicity ratio and the neutron-proton differential
collective transverse flow were firstly pointed out in
Ref.\ \cite{Li:2002qx}. While the threshold production of pions and
kaons and the difference between neutron and proton elliptic flows
were stressed in Refs.\ \cite{Baran:2004ih,Ferini:2006je}. In our previous
investigations, we found that the $\Sigma^-/\Sigma^+$ ratio
\cite{LI:2005zi} at threshold, the $\pi^+-\pi^-$ elliptic flow
difference at moderate transverse momentum \cite{Li:2005gfa}, and the
$\Delta^-/\Delta^{++}$ ratios at large transverse momenta
\cite{Li:2005zz} are sensitive to the density dependence of the
symmetry potential at high densities. Despite of the difficulties of
detecting and analyzing $\Delta$ resonances, the advantage of taking
$\Delta$ as a probe for symmetry potential at high densities is also
obvious: firstly, it is known that most of $\Delta$s can be produced
from the high density region. Secondly, the $\Delta\leftrightarrows
N\pi$ loop sustains the $\Delta$ matter for a relatively long time,
hence the evolution of $\Delta$s should be heavily influenced by the
mean field. Finally, since some of nucleon- and the pion-related quantities
are predicted to be sensitive to the symmetry potential, it is believed that the corresponding
$\Delta$-related quantities will show strong effect on
symmetry potential as well. Thus we are interested in checking one crucial
question: can the transverse momentum distribution of the
$\Delta^-/\Delta^+$ ratio, or more realistically,
the kinetic energy distribution of the $\Delta^0/\Delta^+$ ratio (since it is not easy to reconstruct $\Delta^-$ and its momentum components
through its neutron and $\pi^-$ daughters),  be {\it experimentally} taken as a
sensitive candidate for detecting the symmetry potential at
supranormal densities?
Besides, we found that the transverse momentum dependence of the
neutron/proton ($n/p$) ratio of free nucleons is also sensitive to symmetry
potential \cite{Li:2005kq}. However, since the effect of symmetry
energy term is secondary when comparing it with the
isospin-independent term, one should pay attention to the late stage of HICs
in any transport-model related calculations. In this
work, we also intend to test the influence of different criteria of
constructing clusters on the $n/p$ ratio of free nucleons.

The paper is arranged as follows: in section 2, the UrQMD model is
introduced briefly and the projectile-target combinations to be
investigated are chosen. Section 3 explains different criteria of
constructing free nucleons/fragments and tracing back $\Delta$ resonances at
freeze-out. The effect of the reconstruction criteria on the multiplicities of nucleons and $\Delta$s is discussed.
In section 4, the influence of different reconstruction criteria on the transverse momentum
dependence of the $n/p$ ratio and the kinetic energy dependence of the $\Delta^0/\Delta^{++}$
ratio is illustrated and discussed. The conclusions are given in section 5.

\section{system determination}

The UrQMD transport model \cite{Bass98,Bleicher99} has been
updated for the investigations of intermediate energy HICs.
For details of the updated version of the UrQMD model, the
reader is referred to Ref.\ \cite{Li:2006wc} and
references therein. In this work, the soft EoS with a momentum dependent term
(SM-EoS) is adopted. Two types of density
dependent symmetry potential energy are selected: (1) $F(u)=
u^\gamma$, where $\gamma$ factor varies from $0.5$ (dubbed as
``g05'', soft), $1.0$  (``g1'', moderate) to $1.5$ (``g15'', hard). (2) the so-called
``DDH$\rho^*$'' (very soft) symmetry potential energy from the relativistic
mean-field calculation \cite{Gaitanos:2004cp}. Furthermore, The
''Dirac''-type medium modification of nucleon-nucleon elastic cross
sections discussed in \cite{Li:2006ez} is also considered here.

The choice of the reaction system should be with
great care as well. For $\Delta$-related probes, first of all, in order
to excite enough $\Delta$(1232) resonances but not to excite too many
other higher-lying resonances, the HICs at moderate
beam energies such as $400\sim1000$A MeV should be chosen. At the same time, considering of the
large background spectrum of non-correlated (p,$\pi$) pairs which is
induced especially by heavy system collision \cite{Eskef:2001qg}, a
midsize and neutron-rich projectile-target system is
necessary in order also to obtain an initially large isospin asymmetry
$\delta$ value. Further, inspired by the experiments with 2 or 4 sets of
projectile-target systems of different asymmetry
\cite{Rami:1999xq,Famiano:2006rb} and the model-based investigations
\cite{Li:2005by,Li:2006wc} in order to cancel other potential
effects which are not (obviously) related to isospin, we
select the central ($b=0-2$ fm) $^{132}Sn+^{132}Sn$ and
$^{112}Sn+^{112}Sn$ reactions at $E_b=800$A MeV as examples. While
for nucleon-related probes, the HICs (besides the reactions with
respect to Sn-isotopes we select the Pb+Pb system as well) at a lower beam energy
such as $400$A MeV are preferred and adopted in this work.

\section{Different criteria of constructing clusters and tracing back $\Delta$ resonances}

\begin{figure}
\includegraphics[angle=0,width=0.8\textwidth]{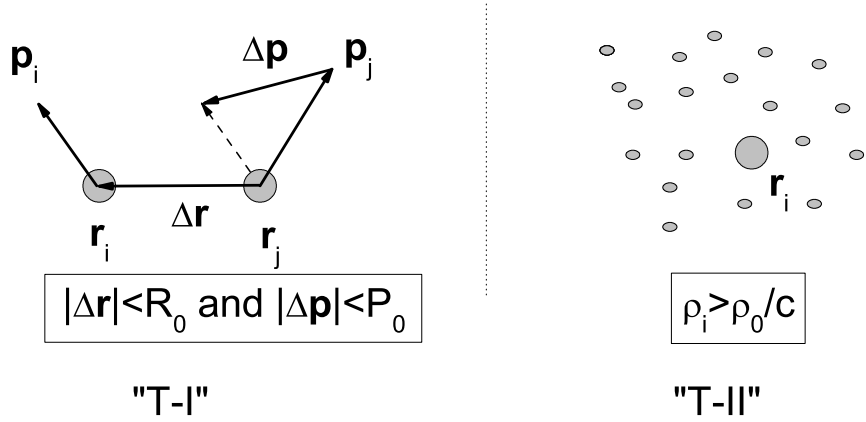}
\caption{Two methods for constructing clusters. Left plot: the
relative distance of two particles $|\Delta\bf{r}|<R_0$ and the
relative momentum of two particles $|\Delta\bf{p}|<P_0$ (``T-I''),
$R_0$ and $P_0$ are free parameters in units of fm and GeV$/c$.
Right plot: coordinate density cut for the particle $i$ in fragments:
$\rho_i>\rho_c=\rho_0/c$ (``T-II''). } \label{fig1}
\end{figure}

\begin{figure}
\includegraphics[angle=0,width=0.7\textwidth]{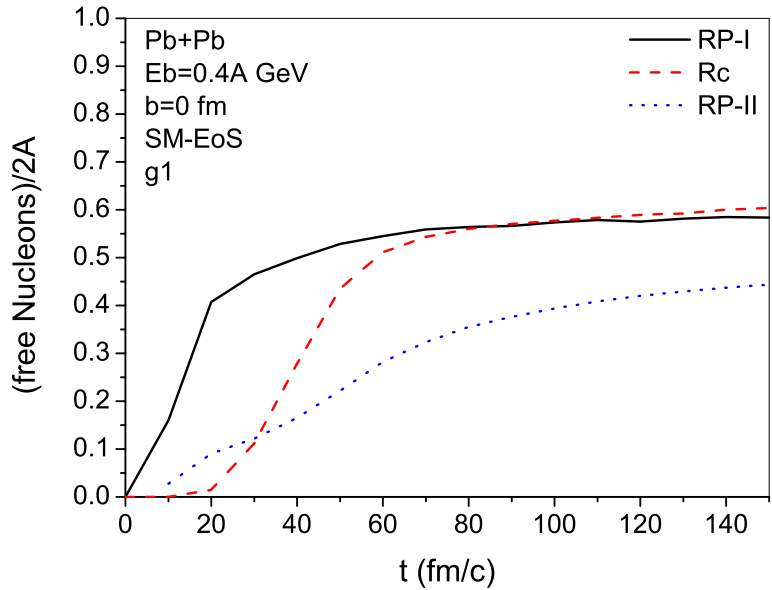}
\caption{Time evolution of the fraction of ``free'' nucleons from Pb+Pb
central collisions at $E_b=0.4$A GeV. The SM-EoS with a linear
symmetry potential is adopted in calculations.
No other physical cuts are selected.
Results of ``finding'' free nucleons from different criteria (``RP-I'', ``RP-II'', and ``Rc'')
are compared (see context).} \label{fig2}
\end{figure}

\begin{figure}
\includegraphics[angle=0,width=0.7\textwidth]{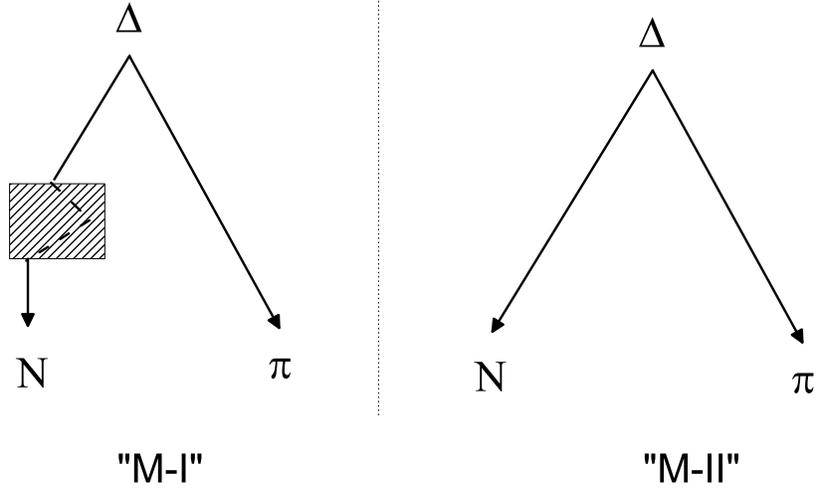}
\caption{Two methods for reconstructing $\Delta$ resonance at
freeze-out. Left plot: the pion produced from a $\Delta$ decay is
found and the pion does not rescatter any more. It is not concerned
if the other nucleon daughter collides further with other particles or not
(marked with a shadowed box). This mode is called as ``M-I''. Right
plot: only the $\Delta$s whose decay daughters do not rescatter any
more are selected. This mode is called as ``M-II''.} \label{fig3}
\end{figure}

\begin{figure}
\includegraphics[angle=0,width=0.8\textwidth]{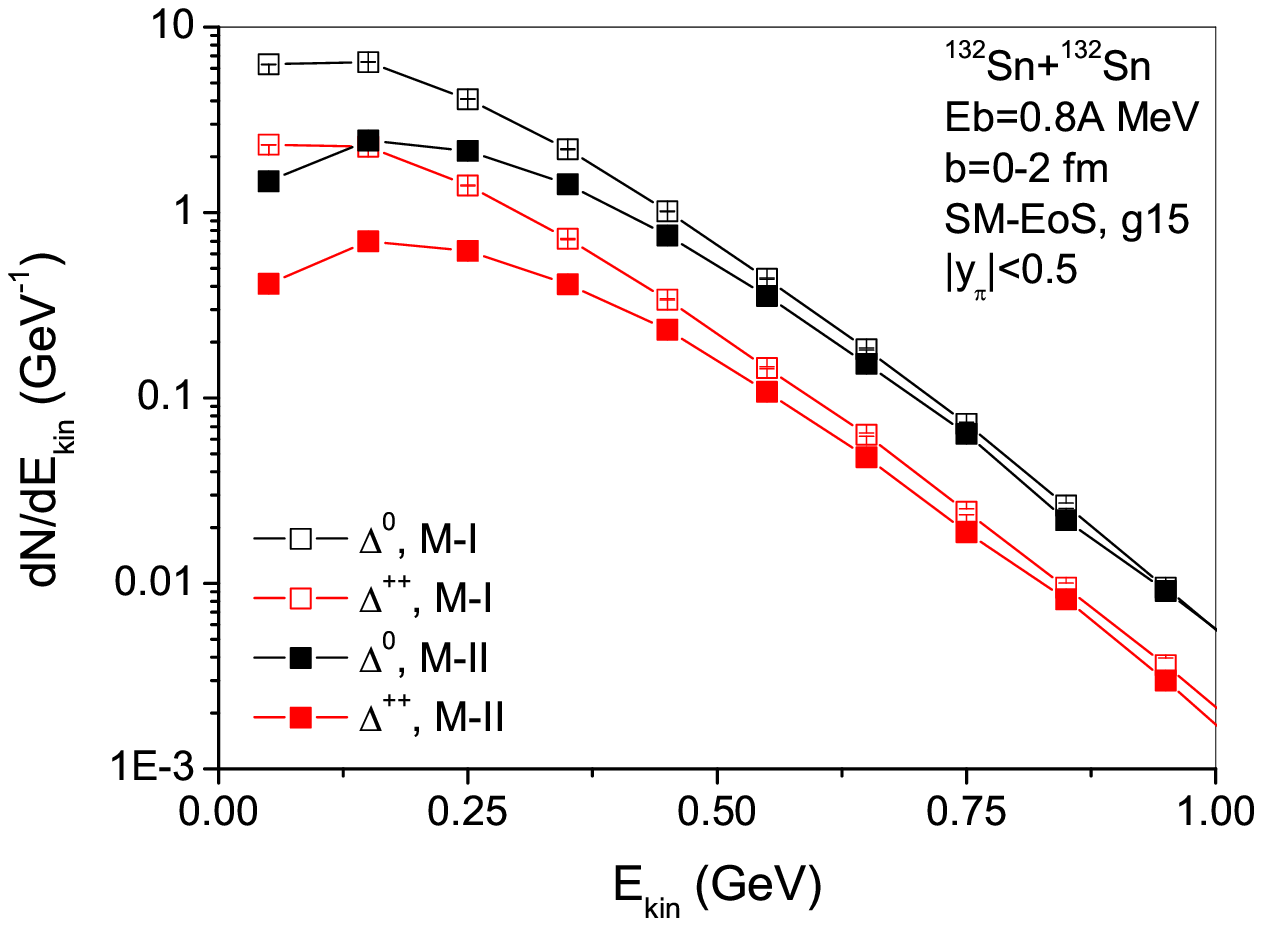}
\caption{Kinetic energy distribution of $\Delta^0$ and $\Delta^{++}$
resonances at freeze-out which is ``found'' by the two methods depicted in
Fig.\ \ref{fig3} from central $^{132}Sn+^{132}Sn$ collisions at
$E_b=0.8$A GeV. The SM-EoS with a stiff symmetry potential
(``g15'') is adopted. The rapidity cut of pions
$|y_{\pi}|<0.5$ is used. } \label{fig4}
\end{figure}

Generally speaking, two methods for constructing clusters exist in the
transport model calculations. The one is called as the coalescence
model, which is normally used in the QMD-like analysis. In the
coalescence model nucleons with relative distances $|\Delta\bf{r}|<R_0$ and
relative momenta $|\Delta\bf{p}|<P_0$ are considered to
belong to one cluster (depicted in the left plot of Fig.\ \ref{fig1}, titled as
``T-I''), otherwise, the nucleons are free. The other one adopts a coordinate density cut $\rho_c$ (depicted
in the right plot of Fig.\ \ref{fig1}, titled as ``T-II'') to separate
''free'' nucleons from other fragments. This method is often used
in the BUU-like model analysis, in which the density of
each real particle is obtained with respect to all of its test
particles. While in QMD-like transport models, each particle is
represent by a Gaussian wave packet in the phase space,
the density of each particle in coordinate space can be expressed
as $\rho_i=\int\rho(\bf{r}_i) \rho d\bf{r}=\int\rho(\bf{r}_i)
\sum_j\rho(\bf{r}_j) d\bf{r}=\frac{1}{(4\pi
L)^{3/2}}\sum_{j}e^{-\frac{(\bf{r}_i-\bf{r}_j)^2}{4 L}}$. In this
work, we examine the freeze-out condition $\rho_i<\rho_c=\rho_0/10$
(dubbed as ``Rc'') to ``find'' free nucleons. Fig.\ \ref{fig2} shows the
time evolution of the fraction of nucleons to be free from Pb+Pb
central collisions at $E_b=0.4$A GeV (by averaging $100$ events).
The SM-EoS with a linear symmetry potential (``g1'') is adopted as an example.
In the ``T-I'' mode, the free nucleons are found out with two sets of
$R_0$ and $P_0$ parameters: (1) ``RP-I'': $R_0=2.8$ fm and $P_0=0.2$
GeV$/c$, (2) ``RP-II'': $R_0=3.5$ fm and $P_0=0.3$ GeV$/c$. In the
``T-II'' mode, the ``Rc'' parameter set is used. It is easy to understand
that with the increase of $R_0$ and $P_0$ values, the number of free
nucleons becomes less and less. Meanwhile, it is seen that the
number of free nucleons with ``RP-I'' mode is almost same as that with
``Rc'' mode after $t\sim 80$fm$/c$ and not sensitive to the time
evolution any more. Before this time, much more nucleons in ``RP-I''
mode than in ``Rc'' mode are taken as free due to the large momentum
difference but the relatively small distance difference  between
nucleons. Therefore, the choice of the cut-time $t_c$ to stop the
transport program will affect the final freeze-out of nucleons if a
too short time is selected. From now on, the $t_c=100$fm$/c$ is
selected for further investigations.

In order to track the resonances at freeze-out, one needs to
look through the detailed history of the time evolution of HICs. In the
UrQMD model there is a standard OSCAR-formatted output file (``ftn20'') which includes
complete event history, that is, besides the
initial and the final freeze-out states, all binary collisions, string-fragmentations, and hadronic
decays of particles are recorded into
this file. From this file, one can easily find the $\Delta$ decay
channel $\Delta\rightarrow N\pi$. It is known that the produced daughters nucleon and pion from this channel
will probably collide further with other particles \cite{Bass:1995pj}.
Thus it is not easy to detect $\Delta$ resonances at freeze-out time experimentally. Currently
there exist two methods in experiments which are employed to ``find'' baryon
resonances: (1) the one-pion $p_t$ distribution is defolded to yield
the mass distribution; (2) the correlated proton and charged pion
pairs are analyzed to yield the invariant mass distribution
\cite{Eskef:2001qg}. In Fig.\ \ref{fig3} we also show two methods
for ''finding'' $\Delta$ resonance at freeze-out: (1) in ``M-I'', the $\Delta$ in which decay channel the
pion daughter will not further
rescatter is found out while, it is not
concerned if the other nucleon daughter rescatters or not. This mode can be taken as a
``contaminated'' reconstructing mode. (2) in ``M-II'', both the pion
and the nucleon from the decay of the $\Delta$ resonance at
freeze-out are confirmed, the $\Delta$ resonance is thus taken to be
reconstructible. And this mode is a ``clean'' reconstructing mode.
Fig.\ \ref{fig4} illustrates the kinetic energy $E_{kin}$ distribution of
$\Delta^0$ and $\Delta^{++}$ resonances at freeze-out ``detected'' by
the above two methods for central $^{132}Sn+^{132}Sn$ collisions at
$E_b=0.8$A GeV. The SM-EoS with a stiff symmetry potential
(``g15'') is adopted as an example. The rapidity cut of
pions $|y_{\pi}|<0.5$ ($y=\frac{1}{2}\log(\frac{E_{cm}+p_{//}}{E_{cm}-p_{//}})$,
$E_{cm}$ and $p_{//}$ are the energy and longitudinal momentum of
the pion in the center-of-mass system) is used. With the ``M-I'' method the amount of
$\Delta$s at small kinetic energies is much larger than that with the
``M-II'' method, which is due to the high chance of
rescattering of the produced nucleon with others. The difference of the $\Delta$ yields with the two methods is
seen to disappear at large $E_{kin}$ due to the fact that the rescattering
probability of the fast nucleons becomes rather rare.

\section{The influence of different reconstruction criteria on sensitive observables}

\begin{figure}
\includegraphics[angle=0,width=0.7\textwidth]{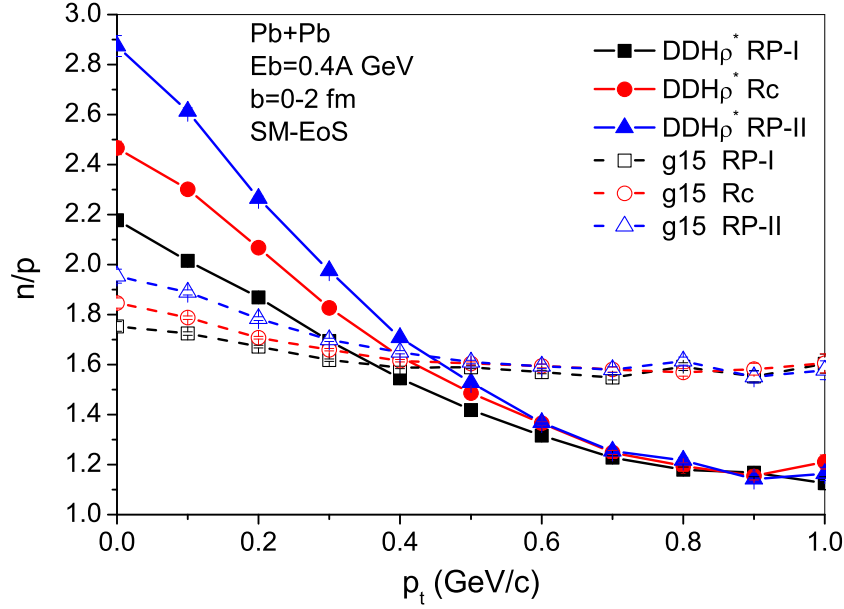}
\caption{Transverse momentum $p_t$ dependence of the $n/p$ ratio for
central Pb+Pb collisions at $E_b=0.4$A GeV. Two sets of symmetry
potential ``g15'' and ``DDH$\rho^*$'' are adopted in calculations. Three
modes ``RP-I'', ``RP-II'', and ``Rc'' are taken into account for
constructing clusters at the cut-time $t_c=100$fm/$c$. } \label{fig5}
\end{figure}

Fig.\ \ref{fig5} shows the transverse momentum $p_t$ dependence of
the $n/p$ ratio of free nucleons from central Pb+Pb collisions at
$E_b=0.4$A GeV. Two sets of density dependent symmetry potential
``g15'' (hard) and ``DDH$\rho^*$'' (very soft) are adopted in calculations. And three modes
``RP-I'', ``RP-II'', and ``Rc'' are taken into account for constructing
clusters at the cut-time $t_c=100$fm$/c$. No rapidity cut is selected. First of all, it is known that the $n/p$ ratio at
low and high transverse momentum reflects the behavior of symmetry
potential at subnormal and supranormal densities, which has been
systematically studied in our previous work \cite{Li:2005kq}, in which the ``RP-II'' is
used in the coalescence-model analysis. With shorter relative phase-space
distances used in the ``RP-I'' mode, more nucleons are counted into free
nucleons, which drives the $n/p$ ratio at small $p_t$ to approaching
the initial $n/p$ ratio of the system (=1.54). Surprisingly,
although the total amount of free nucleons from ``RP-I'' mode is seen
almost same as that from ``Rc'' mode, the $n/p$ ratio of free nucleons
is visibly different from each other at small $p_t$: the ``Rc'' mode
gives a larger $n/p$ ratio. It implies that the two methods of
constructing clusters used for transport models are not equivalent
at small $p_t$ and bring more uncertainties when one hopes to
evaluate the sensitivity of nucleon-related probes to the
symmetry potential. One also finds that the softer the symmetry
potential is, the more sensitive to the free $n/p$ ratio the methods
of constructing clusters are. It is easy to understand since the
soft symmetry potential has larger value than the stiff one at
subnormal densities, thus the softer
symmetry potential influences the freeze-out of particles more
obviously at the late stage. Then, let us move on to see the $n/p$ ratio in the large $p_t$ region
($p_t>0.6$GeV$/c$), it is interesting to see that all of the three
constructing modes have almost no different effect on the values of the $n/p$
ratio. It is obviously due to the fact that the free nucleons with
large $p_t$ are dominantly emitted from high densities at early stage and are
influenced very weakly by the late stage. It also shows that the
$n/p$ ratio of free nucleons at large $p_t$ is a (relatively) clean
and sensitive probe to the symmetry potential at high densities.

\begin{figure}
\includegraphics[angle=0,width=0.7\textwidth]{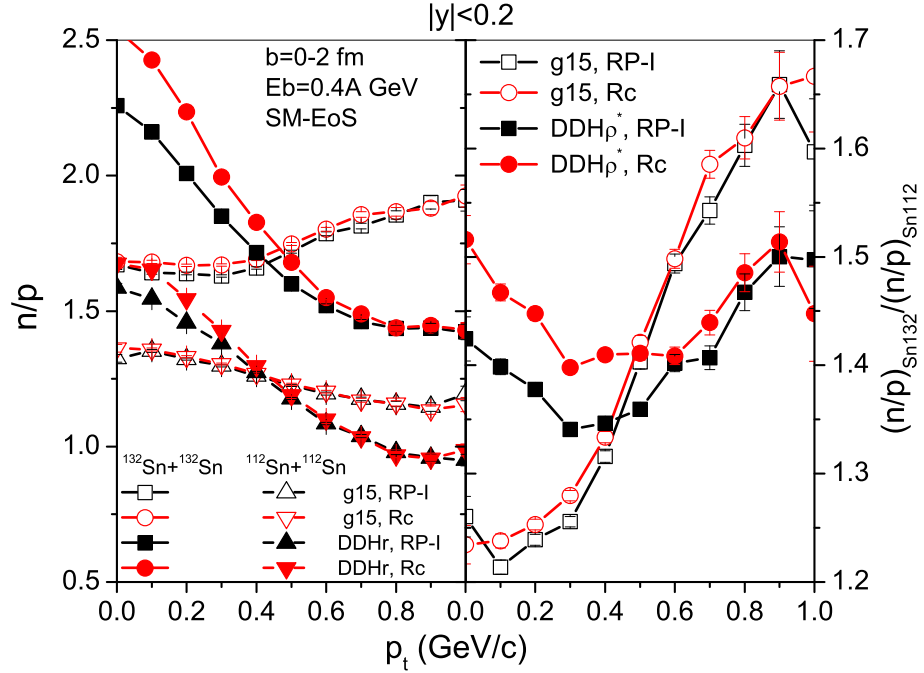}
\caption{Left plot: Transverse momentum $p_t$ dependence of the
$n/p$ ratio of free nucleons for central $^{132}$Sn+$^{132}Sn$ and
$^{112}$Sn+$^{112}Sn$ collisions at $E_b=0.4$A GeV. Two sets of
symmetry potential ``g15'' and ``DDH$\rho^*$'' are adopted in
calculations. Two modes ``RP-I'' and ``Rc'' are taken into account to
construct clusters. The rapidity cut $|y|<0.2$ is chosen. Right
plot: The double ratio between $n/p$ ratios from the
$^{132}$Sn+$^{132}Sn$ system and from the $^{112}$Sn+$^{112}Sn$ system.}
\label{fig6}
\end{figure}

The $p_t$ dependence of the $n/p$ ratio from $^{132}Sn+^{132}Sn$ and $^{112}Sn+^{112}Sn$ systems are also calculated.
The results from the two systems are shown in the left plot of Fig.\ \ref{fig6} separately. In this figure, the
results with modes ``RP-I'' and ``Rc'' are compared and a rapidity cut
$|y|<0.2$ is employed. The results are similar to those in
Fig.\ \ref{fig5}. The effects of the symmetry potential and the
freeze-out criteria on the $n/p$ ratio are week in the more
isospin-symmetric system $^{112}Sn+^{112}Sn$. The right plot of
Fig.\ \ref{fig6} illustrates the double ratio expressed as the
$n/p$ ratio from $^{132}Sn+^{132}Sn$ system divided by the $n/p$ ratio from $^{112}Sn+^{112}Sn$ system,
and is dubbed as $(n/p)_{Sn132}/(n/p)_{Sn112}$. Similar
to the single $n/p$ ratio shown in the left plot of Fig.\ \ref{fig6}, the
effect of freeze-out criteria is still seen at small $p_t$,
especially for the soft symmetry potential case, while at $p_t >
0.6$GeV$/c$, it almost disappears. Therefore, both the single and
the double ratios at large transverse momenta are visibly affected by
the density dependence of the symmetry potential
but invisibly influenced by the different criteria of constructing
clusters and can be taken as good observables for symmetry
potential at high densities.

\begin{figure}
\includegraphics[angle=0,width=0.7\textwidth]{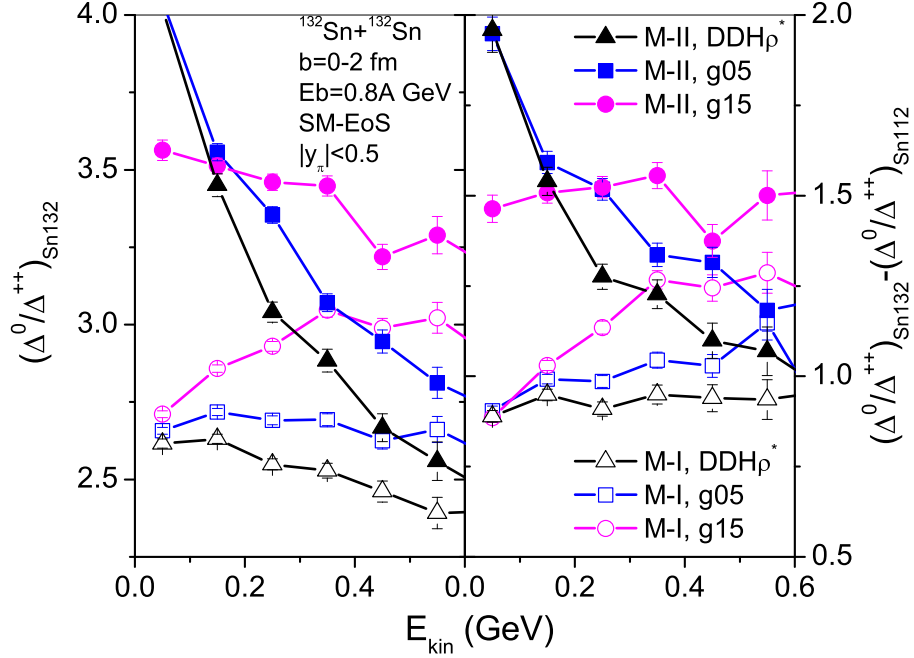}
\caption{Left plot: Kinetic energy distribution of the
$\Delta^0/\Delta^{++}$ ratio at freeze-out. The $\Delta$s are
determined with two methods ``M-I'' and ``M-II'' (see context). The central
$^{132}$Sn+$^{132}Sn$ collisions at $E_b=0.8$A GeV are calculated.
The rapidity cut of pions $|y_{\pi}|<0.5$ is chosen.
Right plot: The $\Delta^0/\Delta^{++}$ ratio from
$^{132}$Sn+$^{132}Sn$ is scaled by the $\Delta^0/\Delta^{++}$ ratio from
$^{112}$Sn+$^{112}Sn$. } \label{fig7}
\end{figure}

Let us finally check the effect of the different analyzing methods
of reconstructing $\Delta$ resonances at freeze-out on its
particle ratio $\Delta^0/\Delta^{++}$. In Fig.\ \ref{fig7} (left
plot) we present the kinetic energy ($E_{kin}$) distribution  of the
$\Delta^0/\Delta^{++}$ ratio at freeze-out where $\Delta$s are
collected with the methods ``M-I'' and ``M-II'' introduced above.
The symmetry potentials ''g05'', ''g15'', and ``DDH$\rho^*$'' are adopted. For each case
$0.36$ million central $^{132}$Sn+$^{132}Sn$ collisions at
$E_b=0.8$A GeV are calculated. And a rapidity cut of pions $|y_{\pi}|<0.5$ is adopted as well.
First of all, the ``crossing'' behavior of the
ratios with soft and stiff symmetry potentials can hardly be seen from the ``M-I'' reconstruction mode.
It implies that the $\Delta^0/\Delta^{++}$ ratio versus $E_{kin}$ is sensitive to the density dependent symmetry potential at high densities.
However, in the ``clean'' ``M-II'' mode, the ``crossing'' behavior re-occurs which means the symmetry potential at low densities shows its role.
Actually, the two modes give a quite different
$\Delta^0/\Delta^{++}$ ratio especially at the small $E_{kin}$. Even at
the large $E_{kin}$ ($>0.5$GeV), the difference does not disappear.
The ``clean'' mode ``M-II'', which implies that the selected $\Delta$s come from more
outer space where densities are low, gives a much large ratio at
all $E_{kin}$ due to the fact that these $\Delta$s are produced from
more neutron-rich region. Even in the scaled ratio by the ratio from $^{112}$Sn+$^{112}Sn$ system (dubbed as
$(\Delta^0/\Delta^{++})_{Sn132}-(\Delta^0/\Delta^{++})_{Sn112}$) shown in the right plot of Fig.\
\ref{fig7}, the huge effect of different reconstructing criteria on
the scaled ratio still exists. Although the
$\Delta^0/\Delta^{++}$ ratio indeed shows its sensitivity to the
density dependence of symmetry potential, the detailed process of the reconstruction
for $\Delta$s at freeze-out should be paid much more attention.

In summary, different criteria of constructing free nucleons and
tracing back $\Delta$ resonances from the intermediate-energy
neutron-rich HICs are discussed with the help of the UrQMD model. It
is found that both the parametrization of relative phase-space
distances used in the coalescence model and the coordinate density
cut for free nucleons modify visibly the single and the double $n/p$
ratios at small transverse momenta, but the influence is weak at
large transverse momenta. Hence, the single and the double $n/p$ ratios
of free nucleons at large transverse momenta can be taken as
sensitive probes for symmetry potential at supranormal densities.
The momentum dependence of the symmetry potential might complicate
this situation and deserves further careful investigation
\cite{Li:2003ts}. Methods of reconstructing $\Delta$ resonances affect the
$\Delta^0/\Delta^{++}$ ratio versus kinetic energy strongly, which
should be paid more attention if one wishes to employ it as a probe
to detect the symmetry potential of the nuclear matter at high
densities.

\section*{Acknowledgments}
We are grateful to W. Trautmann for valuable discussions. We acknowledge support by the Frankfurt Center for
Scientific  Computing (CSC).

\end{document}